\documentclass[reprint,
superscriptaddress,
amsmath,amssymb,aps,showkeys,showpacs,
twoside,final,secnumarabic,
nofootinbib]{revtex4-2}
\usepackage[paperwidth=205mm,paperheight=290mm,top=17mm,bottom=25mm,
inner=17mm,outer=17mm,
twoside]{geometry}

\usepackage[T1,T2A]{fontenc}
\usepackage[utf8]{inputenc}
\usepackage[russian,english]{babel}
\usepackage{color}
\usepackage{graphicx}
\usepackage{dcolumn}
\usepackage{bm} 
\usepackage[unicode=true,colorlinks=true,linkcolor=magenta, urlcolor=blue, citecolor = blue,breaklinks]{hyperref}
\usepackage{multirow}
\usepackage{url}
\usepackage{breakurl}
\DeclareGraphicsExtensions{.eps}

\usepackage{fancyhdr} 
\let\svthefootnote\thefootnote
\newcommand\freefootnote[1]{%
  \let\thefootnote\relax%
  \footnotetext{#1}%
  \let\thefootnote\svthefootnote%
}

\makeatletter 
    
\renewcommand\onecolumngrid{
\do@columngrid{one}{\@ne}%
\def\set@footnotewidth{\onecolumngrid}
\def\footnoterule{\kern-6pt\hrule width 1.5in\kern6pt}%
}

\renewcommand\twocolumngrid{
\def\footnoterule{
\dimen@\skip\footins\divide\dimen@\thr@@
\kern-\dimen@\hrule width.5in\kern\dimen@}
\do@columngrid{mlt}{\tw@}
}%

\makeatother

\begin{document}

\title{Evaluation of EAS directions based on TAIGA HiSCORE data using\\fully connected neural networks} 

\def\addressa{Lomonosov Moscow State University, Skobeltsyn Institute of Nuclear Physics. 
119991 Leninskie gory, 1, bld. 2, Moscow, Russia}
\def\addressb{Research Institute of Applied Physics. 
664003, blv. Gagarina, 20, Irkutsk, Russia}

\author{\firstname{A.P.}~\surname{Kryukov}}
\email[E-mail: ]{kryukov@theory.sinp.msu.ru}
\affiliation{\addressa}
\author{\firstname{S.P.}~\surname{Polyakov}}
\email[E-mail: ]{s.p.polyakov@gmail.com}
\affiliation{\addressa}
\author{\firstname{Yu.Yu.}~\surname{Dubenskaya}}
\affiliation{\addressa}
\author{\firstname{E.O.}~\surname{Gres}}
\affiliation{\addressa, \addressb}
\author{\firstname{E.B.}~\surname{Postnikov}}
\affiliation{\addressa}
\author{\firstname{P.A.}~\surname{Volchugov}}
\affiliation{\addressa}
\author{\firstname{D.P.}~\surname{Zhurov}}
\affiliation{\addressa, \addressb}


\begin{abstract}

The direction of extensive air showers can be used to determine the source of gamma quanta and plays an important role in estimating the energy of the primary particle.
The data from an array of non-imaging Cherenkov detector stations HiSCORE in the TAIGA experiment registering the number of photoelectrons and detection time can be used to estimate the shower direction with high accuracy.
In this work, we use artificial neural networks trained on Monte Carlo-simulated TAIGA HiSCORE data for gamma quanta to obtain shower direction estimates. The neural networks are multilayer perceptrons with skip connections using partial data from several HiSCORE stations as inputs; composite estimates are derived from multiple individual estimates by the neural networks. We apply a two-stage algorithm in which the direction estimates obtained in the first stage are used to transform the input data and refine the estimates. The mean error of the final estimates is less than 0.25 degrees.
The approach will be used for multimodal analysis of the data from several types of detectors used in the TAIGA experiment.
\end{abstract}

\par
\keywords{extensive area shower, EAS direction, Cherenkov detector, machine learning, artificial neural network, multilayer perceptron, skip connections \\[5pt]}

\maketitle
\thispagestyle{empty}
\onecolumngrid

\section{Introduction}\label{intro}

High-energy cosmic rays and gamma quanta colliding with the upper atmosphere produce cascades of secondary particles known as extensive air showers (EASs). These showers can be detected and recorded using a variety of telescopes such as imaging atmospheric Cherenkov telescopes (IACTs), arrays of wide-angle integrating air detectors or water detectors; some experiments such as TAIGA \cite{Budnev22} and LHAASO \cite{Cao22} combine several telescope types.
The data from these observations can be used to identify the primary particle type and estimate its parameters such as energy and direction. 

In this paper, we estimate the EAS direction which is of interest because it can identify the gamma radiation source and is important in estimating the energy of the primary particle. Highly accurate shower direction estimates can be obtained from the timing measurements of multiple detectors spread over a large area such as TAIGA HiSCORE \cite{Gress17}, LHAASO, or HAWC \cite{Abeysekara23}. We use simulated data from TAIGA HiSCORE which is a non-imaging array of wide field-of-view integrating air Cherenkov detector stations.

We use artificial neural networks (ANNs) to obtain shower direction estimates. Convolutional neural networks seem like a natural choice for the problem since the HiSCORE stations are positioned on a grid. However, the previous work using this approach \cite{Vlaskina22, Kryukov23} produced estimates that were significantly less accurate than previously developed methods, e.g. \cite{Tluczykont21}.
In this work, we use multilayer perceptrons with skip connections. They take data from several HiSCORE stations as input and produce partial direction estimates; we combine multiple partial estimates based on different subsets of stations to derive composite estimates. We apply a two-stage algorithm where the process is repeated using two different neural networks. Preliminary direction estimates calculated in the first stage are used to transform the data for the second stage, where the estimates are refined. On average, the resulting estimates are comparable in accuracy to those produced by the conventional approach.

The TAIGA experiment includes other types of detectors colocated with the HiSCORE array. We plan to integrate the proposed approach with analysis of data from other installations, developing multimodal analysis methods.

\section{Methods}\label{methods}

\subsection{Two-stage algorithm}
We use a two-stage algorithm where each stage has its own neural network. The first stage network (ANN-1) calculates preliminary direction estimates and the second stage network (ANN-2) attempts to refine them. This is achieved by transforming the input data for each event based on the preliminary direction estimate: the HiSCORE stations are projected on a plane orthogonal to the estimated direction, and their signal detection times are respectively adjusted.
Since the shower is symmetrical about its axis, the adjusted times resulting from the accurate direction estimate should exhibit symmetry around the intersection point of the shower and the projection plane. We anticipated that input data transformed using moderately inaccurate estimates would allow a neural network to derive the corrections.

\subsection{Input data}
The dataset we use consists of events simulated with the Monte Carlo software for the TAIGA experiment \cite{Postnikov19, Grinyuk20}. The primary particles are gamma quanta with the energy ranging from 100 to 1000 TeV. The range of the zenith and azimuth angles of the EASs is $30^\circ \leq \theta_0 \leq 40^\circ$, $120^\circ \leq \phi_0 \leq 240^\circ$, respectively. Cherenkov radiation produced by the showers is recorded by an array of 121 HiSCORE detector stations with the trigger threshold amplitude of 100 photoelectrons. Some stations data were discarded to ensure that for any event, the interval between the signal detection times by different stations never exceeds 3 $\mu$s. Events with fewer than 10 triggered stations were excluded, except for the training set for ANN-1 that was extended using events with 8 or 9 triggered stations. In the remaining 79745 events, the mean number of the triggered stations is 42.1. These events were randomly divided into training, validation, and test sets, comprising 66490, 7447, and 5808 events, respectively.

The input data for the neural networks are derived from subsets of triggered stations of a fixed size $K$. Each input vector includes the coordinates, the number of detected photoelectrons, and the mean and standard deviation of the detection times for each station in the subset. The stations are sorted by the detection time. The shower direction estimates must be shift invariant, so detection times and coordinates of stations 2 through $K$ are given relative to those of the first station, resulting in $6K-4$ input values. The subset sizes are $K=8$ for ANN-1 and $K=10$ for ANN-2.

We used different algorithms to select station subsets for the training set and for the validation and test sets. For the training set, we aimed to moderately augment the data without introducing excessive redundancy. The subsets were chosen in a way that ensured that no pair shared more than two stations. This resulted in 3.88 million input vectors for ANN-1. For validation and test sets, the direction estimates calculated by the neural network from individual subsets can be combined to produce more accurate composite direction estimates for each event, with accuracy generally improving as the number of subsets increases. For the primary validation and test sets, we selected $m=120$ as the maximum number of subsets per event. Additionally, sets with $m=30$, $m=60$, and $m=240$ were used. 


At the second stage, generation of the training data can follow two approaches: a specialized approach trying to train the neural network to correct the estimation errors made by the specific first stage neural network ANN-1, or a general approach aimed at correcting any estimation errors of the magnitude similar to the errors made by ANN-1. We chose the second approach and randomly generated up to 100 direction estimates for each event with $\theta$ and $\phi$ angles from the normal distributions $\mathcal{N}(\theta_0, 1.5^\circ)$, $\mathcal{N}(\phi_0, 3^\circ)$.
For each pair of an event and a direction estimate we independently selected subsets of triggered stations in a way that ensured that no pair shared more than one station. This resulted in 42.9 million training input vectors. Four pairs of validation and test sets were generated using composite direction estimates calculated in the first stage. Each has the same maximum number of subsets per event $m$ that was used to derive the corresponding composite estimates.

\subsection{Architecture of the neural networks}
\begin{figure*}[h]
\includegraphics[scale=0.5]{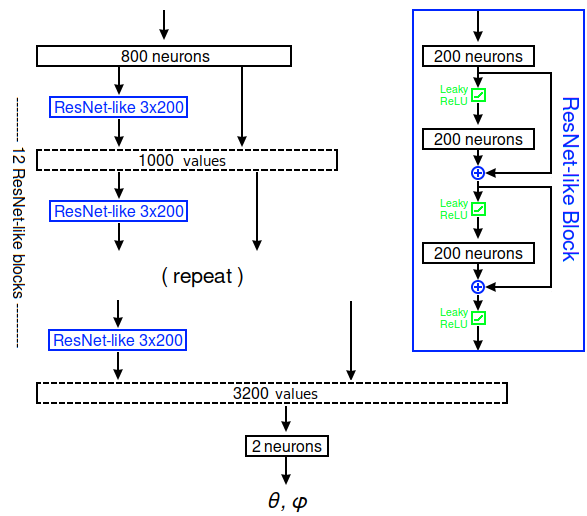}
\caption{\label{fig_architecture} The architecute of the first stage neural network.}
\end{figure*}

The neural network architecture we use employs two types of skip connections: ResNet-like connections using addition and DenseNet-like connections using concatenation \cite{He16a, Huang17}, albeit for fully connected instead of convolutional layers. Smaller blocks with fixed size layers learning residues have their outputs concatenated to previous outputs to form inputs for the next block. Figure \ref{fig_architecture} shows the architecture of ANN-1: the first densely connected layer has 800 neurons, followed by 12 ResNet-like blocks containing 3 layers of 200 neurons with two skip connections. ANN-2 is similar but with 400 neurons in every layer in ResNet-like blocks instead of 200. 
Batch normalization and leaky ReLU activation are applied after each hidden layer. 

The outputs of both neural networks are two values with linear activation which are interpreted as zenith and azimuth angles $\theta_1$, $\phi_1$ for ANN-1 or corrections $\Delta\theta_2$, $\Delta\phi_2$ for ANN-2, in degrees.

\subsection{Training}

The neural networks were implemented and trained using TensorFlow. 
For ANN-1, the loss function we used was $\sin^2 \omega_1$ where $\omega_1$ is the angle between the direction estimate by the neural network and the Monte Carlo direction. The network was trained for 400 epochs, and the weights at epoch 330 were chosen based on the accuracy of the composite estimates for the primary validation set.
For ANN-2, the loss function was $(\Delta\theta_2 - \Delta\theta_0)^2 + \sin^2 \theta (\Delta\phi_2 - \Delta\phi_0)^2$ where $\Delta\theta_0, \Delta\phi_0$ are the corrections that the network needs to learn: $\theta + \Delta\theta_0 = \theta_0$, $\phi + \Delta\phi_0 = \phi_0$ where $(\theta, \phi)$ is the direction estimate that was used to transform the input data. The network was trained for 50 epochs, and the weights at epoch 45 were chosen.
In both cases, Adam optimizer \cite{Kingma14} with the learning rate $10^{-4}$ was used.

\section{Results}\label{results}

The mean value of the angle $\omega_1$ between an EAS direction and a subset-based estimate by ANN-1 for the primary test set is $0.546^\circ$. 
We calculated and compared composite estimates using weighted mean and weighted median values of the subset-based estimates for the primary validation set. 
$1$, $S$, $S^2$, and $S^3$ were used as weights, where $S$ is the total number of photoelectrons detected by the stations of the subset. Table \ref{table1} shows the mean errors of these composite estimates for the primary test set. 
On average, the most accurate composite estimate for the chosen number of subsets was the $S^2$-weighted median estimate.

\begin{table}[h]
\renewcommand{\arraystretch}{1.25}
\renewcommand{\tabcolsep}{3pt}
\begin{center}\caption{Mean angles between EAS directions and their composite estimates by ANN-1.}
\begin{tabular}{|c|c|c|}\hline
             &      mean      &     median     \\
weight       &                &                \\ \hline
$1$          & $0.2802^\circ$ & $0.2654^\circ$ \\ \hline
$S$          & $0.2729^\circ$ & $0.261^\circ$  \\ \hline
$S^2$        & $0.2697^\circ$ & $0.26^\circ$   \\ \hline
$S^3$        & $0.269^\circ$  & $0.2611^\circ$ \\ \hline
\end{tabular}\label{table1}
\end{center}
\end{table}

\begin{figure}[b!]
\includegraphics[scale=0.2]{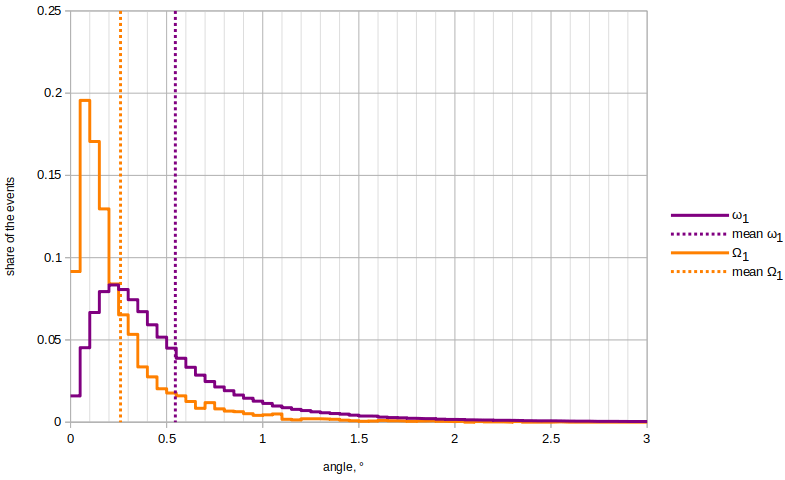}
\caption{\label{fig_omegas1} The distribution of the angles between the EAS direction and its estimates by the first stage neural network. (The part where the fractions are indistinguishable from zero is not shown; $\max \omega_1 = 15.135^\circ$, $\max \Omega_1 = 4.714^\circ$.)}
\end{figure}

The mean value of the angle $\Omega_1$ between an EAS direction and the $S^2$-weighted median composite estimate is $0.26^\circ$. The distribution of $\omega_1$ and $\Omega_1$ angles is shown in Figure \ref{fig_omegas1}.

The mean value of the angle $\omega_2$ between an EAS direction and the composite estimate by ANN-1 with subset-based corrections by ANN-2 is $0.364^\circ$. Mean and median composite corrections with weights $1$, $S$, $S^2$, and $S^3$ were compared for the primary validation set, and the $S^2$-weighted median correction was selected. Table \ref{table2} shows the mean errors of the estimates with these composite corrections for the primary test set.

\begin{table}[h!]
\renewcommand{\arraystretch}{1.25}
\renewcommand{\tabcolsep}{3pt}
\begin{center}\caption{Mean angles between EAS directions and their composite $S^2$-weighted median estimates by ANN-1 with composite corrections by ANN-2.}
\begin{tabular}{|c|c|c|}\hline
             &      mean      &     median     \\
weight       &                &                \\ \hline
$1$          & $0.2235^\circ$ & $0.2193^\circ$ \\ \hline
$S$          & $0.2196^\circ$ & $0.2165^\circ$ \\ \hline
$S^2$        & $0.2176^\circ$ & $0.2153^\circ$ \\ \hline
$S^3$        & $0.2171^\circ$ & $0.2159^\circ$ \\ \hline
\end{tabular}\label{table2}
\end{center}
\end{table}

\begin{figure}[b!]
\includegraphics[scale=0.2]{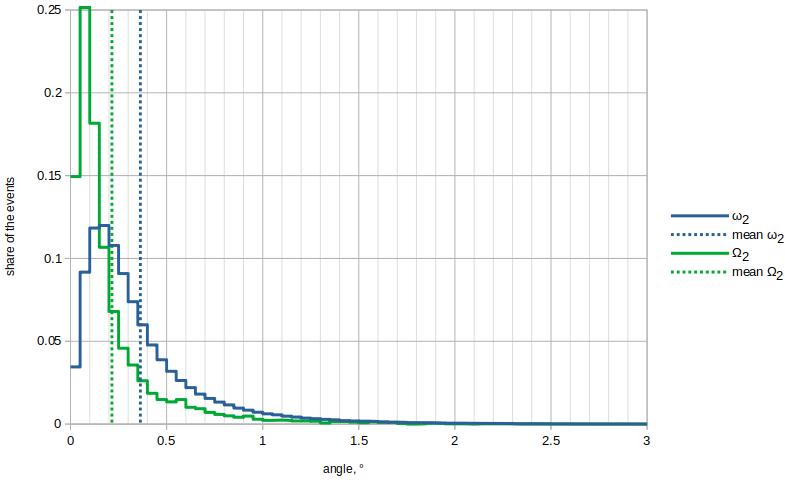}
\caption{\label{fig_omegas2} The distribution of the angles between the EAS direction and its estimates by the second stage neural network. (The part where the fractions are indistinguishable from zero is not shown; $\max \omega_2 = 6.312^\circ$, $\max \Omega_2 = 4.544^\circ$.)}
\end{figure}

The mean value of the angle $\Omega_2$ between an EAS direction and the composite estimate with $S^2$-weighted median composite correction is $0.215^\circ$, median value $0.125^\circ$, RMS $0.344^\circ$. 
The distribution of the $\omega_2$ and $\Omega_2$ angles is shown in Figure \ref{fig_omegas2}.

\begin{figure}[h]
\includegraphics[scale=0.2]{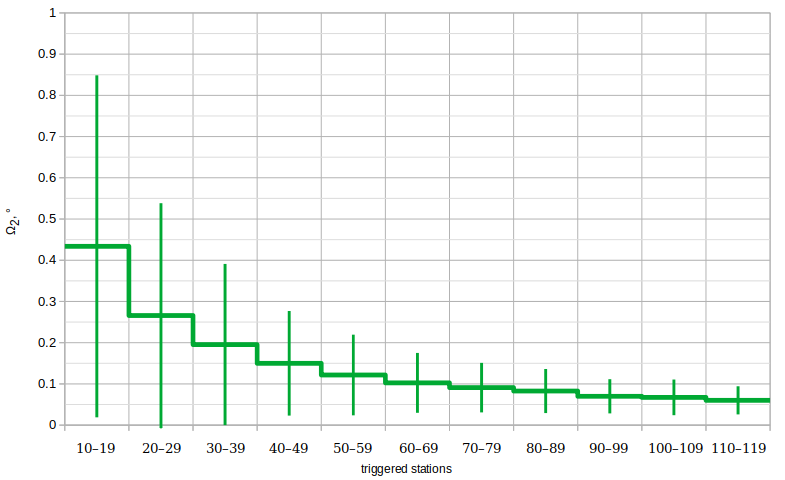}
\caption{\label{fig_Omegas2_vs_n_stations} The mean estimation errors for events grouped by the number of triggered stations.}
\end{figure}

The estimates tend to become more accurate with the number of triggered stations: for the 5237 events with at least 15 triggered stations $\overline{\Omega}_2 = 0.183^\circ$, for the 4670 events with at least 20 triggered stations $\overline{\Omega}_2 = 0.162^\circ$, for the half of the events with at least 38 triggered stations $\overline{\Omega}_2 = 0.116^\circ$, and for the 1341 events with at least 60 triggered stations $\overline{\Omega}_2 = 0.0885^\circ$.
Figure \ref{fig_Omegas2_vs_n_stations} shows the mean estimation errors $\overline{\Omega}_2$ depending on the number of triggered stations.

Less accurate estimates can be obtained at lower computational cost by decreasing the number of station subsets. Conversely, increasing the number of subsets tends to increase the accuracy of the composite estimates. Table \ref{table3} shows the mean errors of the composite estimates obtained from the test sets generated with at most $m=30$, 60, 120, and 240 station subsets per event. All composite estimates use weighted medians with the weights shown in the table.

\begin{table}[h!]
\renewcommand{\arraystretch}{1.25}
\renewcommand{\tabcolsep}{3pt}
\begin{center}\caption{The weights used in composite estimates and corrections and the mean estimation errors by the first and the second stages for the test sets with at most $m$ input entries per event.}
\begin{tabular}{|c|c|c|c|c|}\hline
$m$          & \begin{tabular}{@{}c@{}} stage 1 \\
                weight \end{tabular}
                       & $\overline{\Omega}_1$ & \begin{tabular}{@{}c@{}} stage 2 \\
                                                  weight \end{tabular}
                                                         & $\overline{\Omega}_2$ \\
30           & $S$     & $0.2711^\circ$        & $S$     & $0.2222^\circ$        \\ \hline
60           & $S$     & $0.2654^\circ$        & $S^2$   & $0.2175^\circ$        \\ \hline
120          & $S^2$   & $0.26^\circ$          & $S^2$   & $0.2153^\circ$        \\ \hline
240          & $S^3$   & $0.2561^\circ$        & $S^2$   & $0.2135^\circ$        \\ \hline
\end{tabular}\label{table3}
\end{center}
\end{table}


The accuracy of the conventional method for estimating EAS directions based on TAIGA HiSCORE data is on the order of $0.1^\circ$ \cite{Tluczykont17, Tluczykont21}. When applied to the data that we used, the preliminary results were less accurate: the mean estimation error was $0.349^\circ$, median $0.12^\circ$, RMS $0.67^\circ$ \cite{Osipova24}.

\section{Conclusion}\label{conclusion}

We use multilayer perceptrons with skip connections to obtain estimates of extensive air shower directions based on data from subsets of TAIGA HiSCORE detector stations, then combine them to derive composite estimates. The algorithm has two stages: the estimates calculated in the first stage are used to transform the data for the second stage where the estimates are refined.

Our results show that artificial neural networks can estimate extensive air shower directions with accuracy comparable to that of conventional methods: for the events of our test set, the mean estimation error is $0.215^\circ$, and if we only consider the events detected by at least 60 out of 121 stations, the mean error is $0.0885^\circ$.

\begin{acknowledgments}
The authors would like to thank the TAIGA collaboration for support and
data provision. The work was carried out using equipment provided by the
MSU Development Program.
\end{acknowledgments}

\section*{FUNDING}
This study was supported by the Russian Science
Foundation, grant no. 24-11-00136.


\end{document}